\documentclass[preprint,final,5p,times,twocolumn]{elsarticle}
\usepackage{rotating,color,subfigure,amssymb}
\usepackage{alphalph}
\usepackage{amsmath}
\usepackage[T1]{fontenc}
\usepackage{amssymb}
\journal{Physics Letters B}

\begin{document}


\title{First measurement of polarisation transfer $C^n_{x'}$ in deuteron photodisintegration.}

\author[UoY]{M.~Bashkanov}\ead{mikhail.bashkanov@york.ac.uk}
\author[UoY]{D.P.~Watts}
\author[UoR]{S.~Kay}

\author[UBasel]{S.~Abt}
\author[KPM]{P.~Achenbach}
\author[KPM]{P.~Adlarson}

\author[UoB]{F.~Afzal}
\author[UoR]{Z.~Ahmed}
\author[KSU]{C.S.~Akondi}

\author[UoG]{J.R.M.~Annand}
\author[KPM]{H.J.~Arends}
\author[UoB]{R.~Beck}

\author[KPM]{M.~Biroth}
\author[Dubna]{N.~Borisov}
\author[INFN]{A.~Braghieri}
\author[GWU]{W.J.~Briscoe}

\author[KPM]{F.~Cividini}
\author[SMU]{C.~Collicott}
\author[Pavia,INFN]{S.~Costanza}
\author[KPM]{A.~Denig}

\author[GWU]{E.J.~Downie}
\author[Giessen,KPM]{P.~Drexler}
\author[UoY]{S.~Fegan}
\author[Tomsk]{A.~Fix}

\author[UoG]{S.~Gardner}
\author[UBasel]{D.~Ghosal}
\author[UoG]{D.I.~Glazier}

\author[Dubna]{I.~Gorodnov}
\author[KPM]{W.~Gradl}
\author[UBasel]{M.~G\"unther}

\author[INR]{D.~Gurevich}
\author[KPM]{L. Heijkenskj{\"o}ld}

\author[MAU]{D.~Hornidge}
\author[UoR]{G.M.~Huber}
\author[UBasel]{A.~K{\"a}ser}
\author[KPM,Dubna]{V.L.~Kashevarov}

\author[RBI]{M.~Korolija}
\author[UBasel]{B.~Krusche}
\author[Dubna]{A.~Lazarev}

\author[UoG]{K.~Livingston}
\author[UBasel]{S.~Lutterer}
\author[UoG]{I.J.D.~MacGregor}
\author[KSU]{D.M.~Manley}

\author[KPM,MAU]{P.P.~Martel}
\author[UoM]{R.~Miskimen}
\author[KPM]{E.~Mornacchi}

\author[UoG]{C. Mullen}
\author[Dubna]{A.~Neganov}
\author[KPM]{A.~Neiser}

\author[KPM]{M.~Ostrick}
\author[KPM]{P.B.~Otte}
\author[UoR]{D.~Paudyal}

\author[INFN]{P.~Pedroni}
\author[UoG]{A. Powell}
\author[UoC]{S.N.~Prakhov}

\author[KPM]{V.~Sokhoyan}
\author[UoB]{K.~Spieker}
\author[KPM]{O.~Steffen}

\author[GWU]{I.I.~Strakovsky}
\author[UBasel]{T.~Strub}
\author[RBI]{I.~Supek}

\author[UoB]{A.~Thiel}
\author[KPM]{M.~Thiel}
\author[KPM]{A.~Thomas}

\author[Dubna]{Yu.A.~Usov}
\author[KPM]{S.~Wagner}

\author[KPM]{J.~Wettig}
\author[KPM]{M.~Wolfes}
\author[UoY]{N.~Zachariou}

\address[UoY]{Department of Physics, University of York, Heslington, York, Y010 5DD, UK}
\address[UoR]{University of Regina, Regina, SK S4S-0A2 Canada}
\address[KSU]{Kent State University, Kent, Ohio 44242, USA}
\address[UoG]{SUPA School of Physics and Astronomy, University of Glasgow, Glasgow, G12 8QQ, UK}
\address[UBasel]{Department of Physics, University of Basel, Ch-4056 Basel, Switzerland}
\address[KPM]{Institut f\"ur Kernphysik, University of Mainz, D-55099 Mainz, Germany}
\address[UoB]{Helmholtz-Institut f\"ur Strahlen- und Kernphysik, University Bonn, D-53115 Bonn, Germany}

\address[Dubna]{Joint Institute for Nuclear Research, 141980 Dubna, Russia}

\address[INFN]{INFN Sezione di Pavia, I-27100 Pavia, Pavia, Italy}

\address[GWU]{Center for Nuclear Studies, The George Washington University, Washington, DC 20052, USA}
\address[SMU]{Department of Astronomy and Physics, Saint Mary's University, E4L1E6 Halifax, Canada}
\address[Pavia]{Dipartimento di Fisica, Universit\`a di Pavia, I-27100 Pavia, Italy}
\address[Giessen]{II. Physikalisches Institut, University of Giessen, D-35392 Giessen, Germany}
\address[INR]{Institute for Nuclear Research, RU-125047 Moscow, Russia}

\address[MAU]{Mount Allison University, Sackville, New Brunswick E4L1E6, Canada}
\address[RBI]{Rudjer Boskovic Institute, HR-10000 Zagreb, Croatia}

\address[UoM]{University of Massachusetts, Amherst, Massachusetts 01003, USA}
\address[UoC]{University of California Los Angeles, Los Angeles, California 90095-1547, USA}
\address[RIP]{Racah Institute of Physics, Hebrew University of Jerusalem, Jerusalem 91904, Israel}
\address[RU]{Department of Physics and Astronomy, Rutgers University,
Piscataway, New Jersey, 08854-8019}

\address[JLab]{Jefferson Lab, 12000 Jefferson Ave., Newport News, VA 23606, USA}

\address[Tomsk]{Tomsk Polytechnic University, 634034 Tomsk, Russia}

\cortext[coau]{Corresponding author }


\date{\today}

\begin{abstract}
A first measurement of the polarisation transfer from a circularly-polarised photon to the final state neutron ($C^n_{x'}$) in deuterium photodisintegration has been carried out. This quantity is determined over the photon energy range 370~--~700~MeV and for neutron centre-of-mass breakup angles $\sim45-120^{\circ}$. The polarisation of the final state neutrons was determined by an ancillary large-acceptance nucleon polarimeter, surrounding a cryogenic liquid deuterium target within the Crystal Ball detector at MAMI. The polarimeter characterised $(n,p)$ charge exchange of the ejected neutrons to determine their polarisation. The new $C^n_{x'}$ data are also compared to a theoretical model based on nucleonic and nucleon resonance degrees of freedom constrained by the current world-database of deuterium photodisintegration measurements. Structures in $C^n_{x'}$ observed in the region of the $d^*(2380)$ could not be explained by conventional models of deuteron photodisintegration. 

\end{abstract}

\maketitle


\section{\label{sec:Intro} Introduction}
Despite study for over a century~\cite{Chadwick} the photodisintegration of the deuteron, one of the most basic reactions of nuclear physics, has lacked full experimental constraint. Although the cross section is well determined, there is a paucity of measurements of polarisation observables for the photodisintegration process. This issue is being addressed with a new programme of measurements in the A2 collaboration at MAMI to significantly expand the database of polarisation observables. The photon energies available at MAMI (0.15-1.5 GeV) enable the reaction process to be probed at distance scales where both the nucleonic and sub-nucleonic (quarks) substructure of the deuteron play a role. Such studies are of particular current importance as, alongside constraints on the role of conventional nucleon resonances and meson exchange currents, polarisation observables may provide sensitivity to more exotic QCD possibilities such as the six-quark containing (hexaquark) $d^*(2380)$. The $d^*(2380)$  has recently been evidenced in a range of nucleon-nucleon scattering reactions~\cite{mb,MB,MBC,TS1,TS2,MBA,MBE1,MBE2} from which quantum numbers $I(J^P)=0(3^+)$, a mass $M_{d^*}\sim2380$~MeV and width $\Gamma \sim70$~MeV have been derived. In photoreactions this corresponds to a pole at $E_{\gamma}\sim570$~MeV. Constraints from photoreactions on the existence, properties and electromagnetic coupling of the $d^*(2380)$ would have important ramifications for the emerging field of non-standard multiquark states, and potentially for the dynamics of condensed matter systems such as neutron stars~\cite{nstars}.

The deuteron photodisintegration reaction process~\cite{ArenD} can be described by 12 independent complex helicity amplitudes. Achieving full information on these amplitudes requires a measurement programme of unpolarised, single-polarisation and double-polarisation observables in which combinations of photon beam polarisation, deuteron polarisation and final state nucleon polarisations are determined. We discuss the world database of measurements for deuteron photodisintegration in the relevant photon energy range, 0.15-1.5~GeV, below. 

For the energy ranges studied in the current work the cross section for deuteron photodisintegration has been determined over a wide range of kinematics ~\cite{DAPHNE}(A2@MAMI). Recent measurements~\cite{mbMainz}(A2@MAMI) of the single-polarisation observable $\Sigma$, accessed through disintegration by linearly polarised photon beams, have also been obtained.  Measurement of the target polarisation asymmetry (T)~\cite{TargetD} is contrained  by data at INS.  Measurement of the induced recoil nucleon polarisation of the final state neutron, $P^n_y$, has been obtained only recently, and indicated the induced neutron polarisation approaches 100\% in the region where the $d^*$ may be expected to contribute, mirroring features observed for the induced proton polarisation ~\cite{TOK1,TOK2} ($P^p_y$). The behaviour of both $P^n_y$ and $P^p_y$ are not described by available models based on nucleonic degrees of freedom. For double-polarisation observables, there is only a single data point from a measurement of the transferred polarisation to the ejected proton from helicity-polarised photons, $C_{x'}^p$, obtained at a centre-of-mass (CM) breakup angle of $\Theta_p^{CM}\sim90^{\circ}$ at $E_{\gamma}$=475 MeV. 

In this work, we present the first measurement (in any photon energy range) of the transferred polarisation to the neutron in deuteron photodisintegration, $C^n_{x'}$. The measurement was obtained using the Crystal Ball detector in  A2@MAMI, sampling photon energies $E_{\gamma}=370-700$~MeV and CM breakup angles of $\Theta_{n}^{CM}=45-120^{\circ}$. 

The $C^n_{x'}$ data provides new constraints on the fundamental reaction process for deuteron photodisintegration. The data are compared to a theoretical model based on nucleon and nucleon resonance degrees of freedom in a diagrammatic approach, constrained by the current world data base of deuteron photodisintegration data.

\section{Experimental Details}
\begin{figure}[!h]
\begin{center}
\includegraphics[width=0.40\textwidth,angle=0]{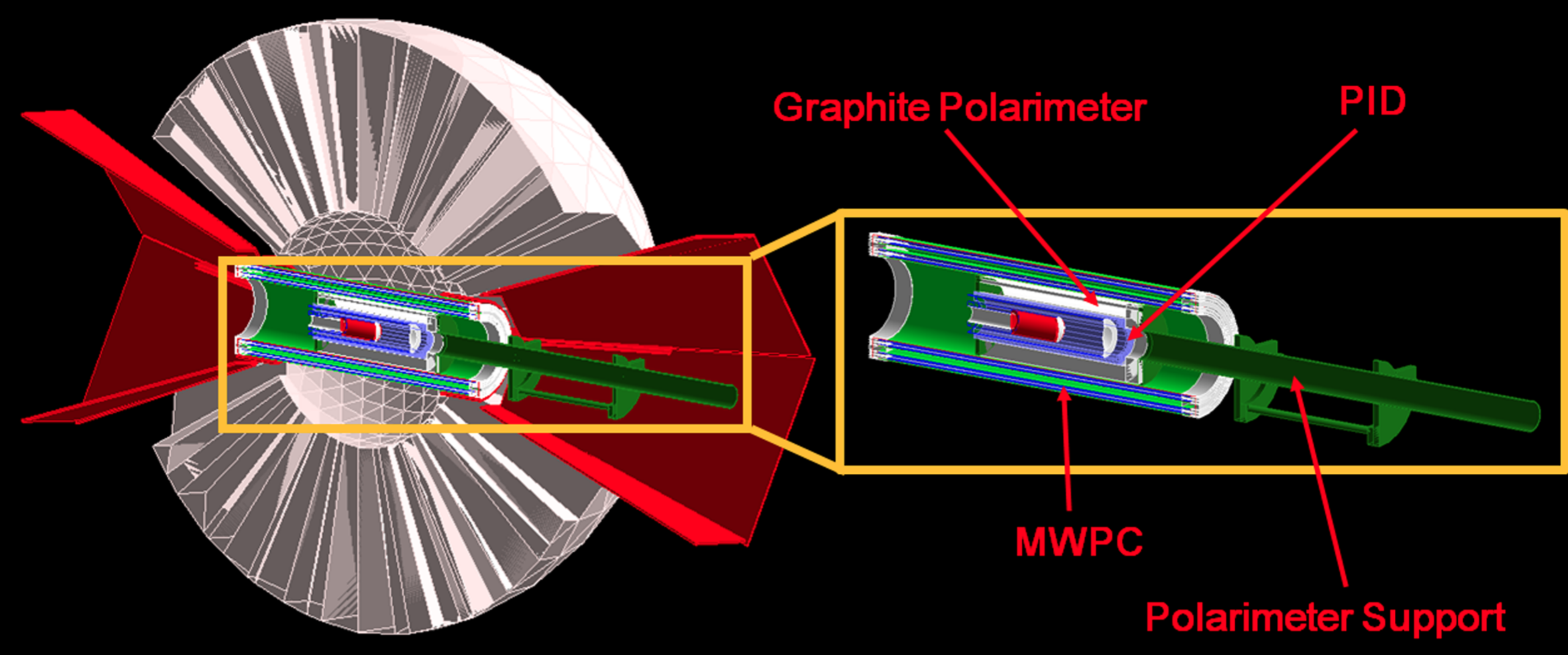}
\end{center}
\caption{Crystal Ball setup during the polarimeter beamtime. The cryogenic target (red cell) is surrounded by the PID barrel (blue), the graphite polarimeter (grey), the MWPC (blue/green) and the Crystal Ball (white).}

\label{Exper}
\end{figure}

The measurement employed a new large acceptance neutron polarimeter~\cite{proposal} within the Crystal Ball detector at the A2@MAMI~\cite{MAMI} facility during a 600 hour beamtime. A 1557~MeV longitudinally-polarised electron beam impinged on either a thin amorphous (cobalt-iron alloy) or crystalline (diamond) radiator, producing circularly (alloy) or elliptically (diamond) polarised bremsstrahlung photons. As linear photon beam polarisation is not used to extract $C_{x'}^{n}$, equal flux from the two linear polarisation settings were combined to increase the circularly-polarised yield~\footnote{The extracted $C_{x'}^{n}$ for the pure circular and combined linear beam data gave consistent results within the statistical accuracy of the data.}. The photons were energy-tagged ($\Delta E\sim2$~MeV) by the Glasgow-Mainz Tagger~\cite{Tagg} and impinged on a 10 cm long liquid deuterium target cell.  Reaction products were detected by the Crystal Ball (CB)~\cite{CB}, a highly segmented NaI(Tl) photon calorimeter covering nearly 96\% of $4\pi$ steradians. For this experiment, a new dedicated 24 element, 7~cm diameter and  30~cm long plastic-scintillator barrel (PID-POL)~\cite{PID} surrounded the target, with a smaller diameter than the earlier PID detector~\cite{PID}, but provided similar particle identification capabilities. A 2.6~cm thick cylinder of analysing material (graphite) for nucleon polarimetry was placed around PID-POL, covering polar angles $\Theta=12-150^{\circ}$ and occupying the space between PID-POL and the Multi Wire Proportional Chamber (MWPC)~\cite{MWPC}. The MWPC provided charged-particle tracking for particles passing out of the graphite into the CB. At forward angles, an additional 2.6~cm thick graphite disc covered the range $\Theta=2-12^{\circ}$~\cite{PID, MBPy}.  The GEANT4 representation of the setup can be seen in Fig.~\ref{Exper}.

The $d(\gamma^{\odot/\otimes},p\vec{n})$ events of interest consist of a primary ejected proton track and a kinematically reconstructed neutron, which undergoes a $(n,p)$ charge-exchange reaction in the graphite to produce a secondary proton which subsequently produces signals in the MWPC and CB. A schematic of the experimental setup is shown in Fig.~\ref{Exper}. The primary proton was identified using the correlation between the energy deposits in the PID and CB using $\Delta E-E$ analysis~\cite{PID} with associated track information obtained from the MWPC. The intercept of the primary proton track with the photon beamline allowed determination of the production vertex, enabling the yield originating from the target cell windows to be removed. Neutron $^{12}$C$(n,p)$ charge exchange candidates required an absence of a PID-POL signal on the reconstructed neutron path into the graphite, in coincidence with a secondary proton track in the MWPC and a corresponding hit in the CB. The reconstructed incident neutron angle ($\Theta_{n}$) was determined kinematically from $E_{\gamma}$ and the production vertex coordinates. A distance of closest approach condition was imposed to ensure crossing of the (reconstructed) neutron track and the secondary proton track. Once candidate proton and neutron tracks were identified, a kinematic fit was employed to increase the sample purity and improve the determination of the reaction kinematics, exploiting the fact that the disintegration can be constrained with measurements of two kinematic quantities while three ($\Theta_{p}, T_{p} $ and $ \Theta_{n}$) are measured in the experiment~\footnote{The fit is constrained taking the photon energy as fixed, the primary proton measured and primary neutron unmeasured.}. A 10\% cut on the probability function was used to select only events from regions where a uniform probability is observed~\cite{mbMainz}.

\section{Determination of spin transfer}

\begin{figure}[!h]
\begin{center}
\includegraphics[width=0.48\textwidth,angle=0]{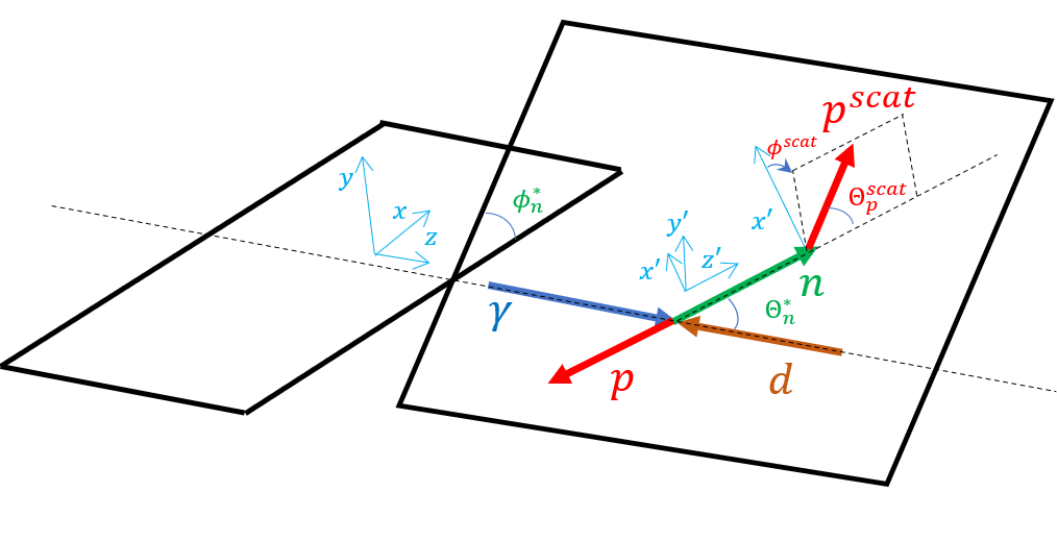}

\end{center}
\caption{Kinematics of the reaction in the centre of mass system. The z-axis is oriented along the beam, y-axis is perpendicular to the ground; $z'$ axis is oriented along the ejectile neutron direction, $y'$ axis is perpendicular to the reaction plane.}
\label{Kin}
\end{figure}

The cross section for deuterium photodisintegration by circularly polarised photons with determination of recoil neutron polarisation is given~\cite{GG} by:

\begin{equation}
\frac{d\sigma}{d\Omega}=\left(\frac{d\sigma}{d\Omega}\right)_0 \cdot[1+C^n_{x'}\cdot P_{\gamma}^{\odot}\cdot A\sin(\phi^{scat}) +P_{y}A\cos(\phi^{scat})],
\end{equation}
where $\left(\frac{d\sigma}{d\Omega}\right)_0$ is the unpolarised cross-section, $C^n_{x'}$ is the transferred polarisation from the photon to the recoiling neutron, $P_{\gamma}^{\odot}$ is the circular polarisation of the incoming photon (which in our case is flipped between positive and negative values) and $A$ is the analysing power for the  $^{12}$C$(n,p)$ reactions occurring in a graphite analysing medium (the polarimeter). $P_{y}$ is the (helicity independent) induced nucleon polarisation. $\phi^{scat}$ is the azimuthal angle of the scattered proton from $^{12}$C$(n,p)$ in the primed frame, where the z-axis is in the direction of the nucleon and the x and y axes are defined relative to the reaction plane (see Fig~\ref{Kin}).

To extract $C_{x'}^{n}$ from the measured data a log-likelihood ansatz was employed. The event-by-event likelihood function, proportional to the event yield (the product of cross section and acceptance), can be defined as:
\begin{equation}
    L_i=c_i\left[ 1+A_{y,i}(C^n_{x'}\cdot P_{\gamma,i}^{\odot}\cdot \sin(\phi^{scat}_i) + P_{y}\cos(\phi^{scat}))\right] A_{i},
\end{equation}
where $i$ is the index of the event under consideration, $c_i$ is a normalization coefficient, $A_{i}$ is the detector acceptance.  $A_{y,i}$ is the analysing power for the $^{12}$C$(n,p)$ reaction for the kinematics of the event ($i$).

The experimental dataset, comprising $i$ events, is fitted by a log-likelihood function, obtained by taking the log of equation 2:
\begin{equation}
    \log L=b+\sum_i \log\left[ 1 +A_{y,i}(C^n_{x'}\cdot P_{\gamma,i}^{\odot}\cdot \sin(\phi^{scat}_i + P_{y}\cos(\phi^{scat}))  \right] \label{Eq:LogLike},
\end{equation}
 The summation $(i)$ reflects how the function is minimised by fitting to all data. The fit has free parameters $b$ and $C^n_{x'}$ while $P_{\gamma,i}^{\odot}\cdot A_{y,i}$ are fixed and calculated on an event-by-event basis. $P_{y}$ is helicity independent and was taken from Ref~\cite{MBPy}. The extracted $C^{n}_{x'}$ is rather insensitive to the adopted value of $P_{y}$ giving maximum variation of 0.02. The constant $b$ is an observable-independent constant, which absorbs the normalization coefficient and detector acceptance, but whose contribution cancels in the likelihood extraction of $C^n_{x'}$ (derived from an asymmetry of yields between the two beam helicity states which are flipped regularly with a period of $\sim1s$.
 
 The fitting procedure used unbinned azimuthal scatter distributions to mitigate any bin-size dependent systematic effects. The variation of $C_{x'}$ with photon energy and neutron angle is assumed continuous and parameterised as a smooth function
\begin{equation}
    C_{x'}=\sum_{i=1}^{L_{max}} a_i\cdot P^i_1(\Theta_n^*),
\end{equation}
where $P^i_1$ are associated Legendre functions of the first order and $a_i$ are smooth energy dependent functions\footnote{In this particular case smooth functions were parameterised by equidistant(50~MeV apart) Gaussians with fixed 25~MeV $\sigma$ and arbitrary strength. To avoid biases, the central values of Gaussians were randomised for each bootstrap cycle}. In our case we used $L_{max}=5$. The results are largely insensitive to this choice, so long as the function covers the full parameter space (as is the case here). If $L_{max}$ exceeds the degrees of freedom for the  $C_{x'}$ description, the decomposition coefficients would become correlated, while $C_{x'}$ itself stays unchanged. To ensure accurate calculation of the errors in extraction of $C_{x'}$, where there is potential for correlated errors, we employ a bootstrap technique~\cite{Ale}. From our sample of $N$ events we randomly select $N$ events, allowing repetitions, and make a likelihood fit to extract $C_{x'}$ as a surface function $C_{x'}=f(\Theta,E_{\gamma})$. Multiple repetitions of the procedure provide the most likely $C_{x'}$ along with determination of the associated statistical and systematic errors. 

The fixed parameters in the likelihood fit to the data (equation 3) are $P_{\gamma,i}^{\odot}$ and $A_{y,i}$, which are both determined on an event-by-event basis. $P_{\gamma,i}^{\odot}$ is calculated analytically from the incident electron beam energy and the tagged photon energy~\cite{gPol}. The systematic uncertainty in helicity polarisation from the calculation is estimated~\cite{circPi} to be 3\%. The magnitude of  $A_{y,i}$ depends on the ejectile neutron energy and scattered proton polar angle for the identified $^{12}$C$(n,p)$ reaction. The $A_{y,i}$ for each event was taken from the SAID parameterisation~\cite{SAID} of free n-p scattering,  modified to account for the n-p reactions occuring in $^{12}$C using an experimental determination of $A_y$ for $^{12}$C$(n,p)X$ by JEDI@Juelich~\cite{Jedi}. The magnitude of the SAID analysing powers were calibrated to the JEDI data by the function: $A_y(n^{12}C)/A_y(np)=1+e^{(1.82-0.014E_n[MeV])}$. This modified analysing power function described the JEDI data with a $\chi^{2}$ close to 1. The angular distributions from SAID agreed with the JEDI data. This enhancement originates from  the contribution of coherent nuclear processes, such as $^{12}$C$(n,p)^{12}$N. For the lowest photon energies sampled in the $\Delta$ region the typical neutron analysing power is enhanced over the SAID prediction by $\sim$30\%. The size of the enhancement reduces with increasing photon (neutron) energy e.g. it is below $\sim$5\% for $E_{\gamma}$ above 500 MeV. To avoid regions of low analysing power, events were only retained for analysis if $A_y(np)\geq$0.1 and the proton polar scattering angle relative to the direction of the neutron, $\Theta^{scat}_{p}$, was in the range 15-45$^{\circ}$. The systematic uncertainty of the analysing power determination is derived from the uncertainty of the JEDI $^{12}$C$(n,p)X$  measurement to which it is calibrated (estimated to be 10\% \footnote{The JEDI $^{12}$C$(n,p)$ data are unpublished. However systematics can be obtained from the  published $^{12}$C$(d,d)$ analysing power measurement which used common apparatus and the same beamtime. The systematic uncertainty is shown to be dominated by the uncertainty in beam polarisation and established to be 10\% \cite{Jedi1})}).


Relaxing the analysis cuts (increasing the minimum probability in the kinematic fit up to 40\%~\cite{MBPy}) gave negligible change in the extracted $C_{x'}$ (below 0.02). The consistency of $C_{x'}$ extraction from separated amorphous and diamond radiator datasets gave the dominant contribution to the systematic error budget (typically 0.2 for much of the parameter space). However the magnitude of this error is driven by the available statistics, and could be reduced in future measurements with higher statistics.

\begin{figure}[!h]
\begin{center}
\includegraphics[width=0.40\textwidth,angle=0]{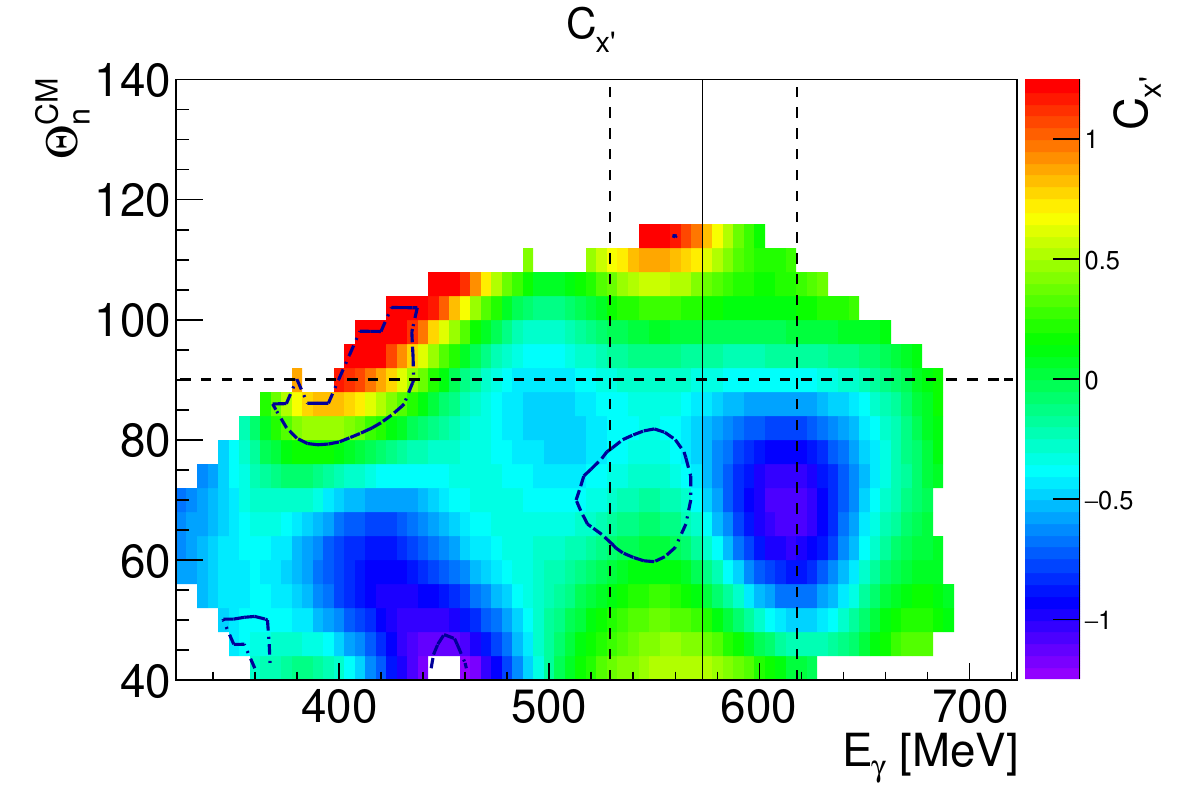}
\includegraphics[width=0.40\textwidth,angle=0]{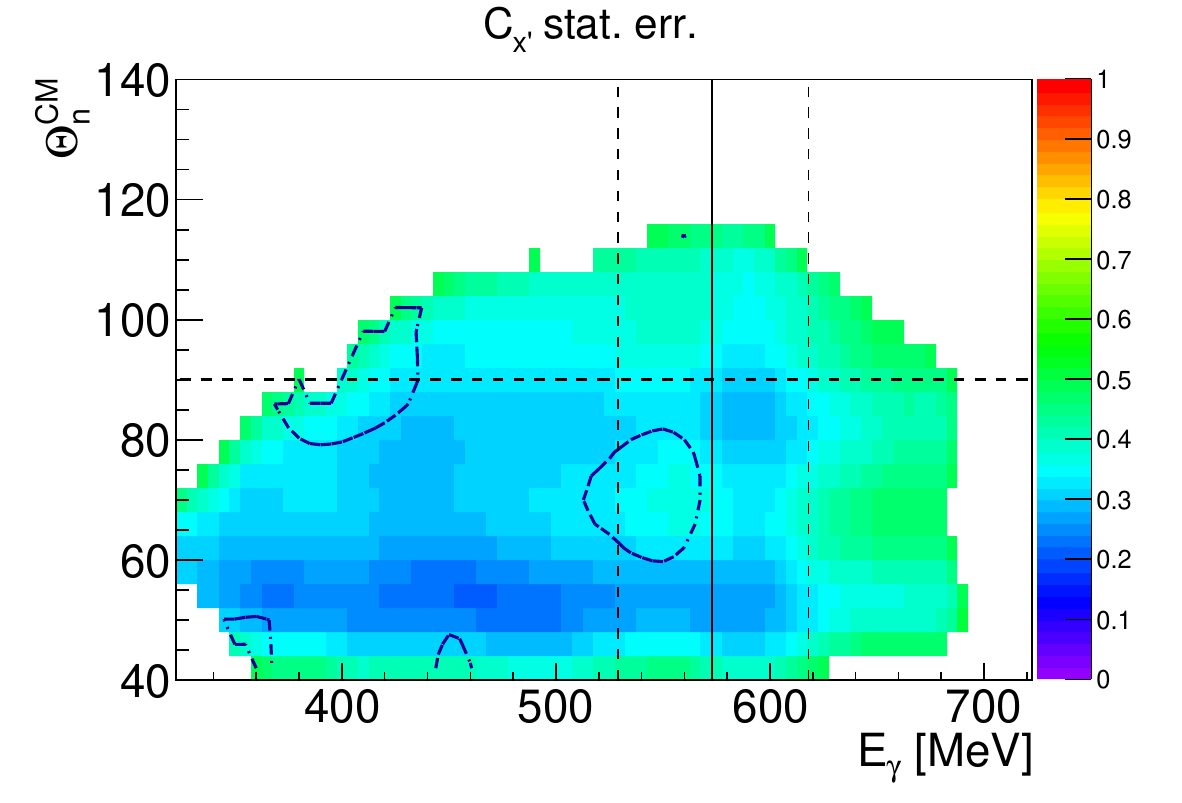}
\includegraphics[width=0.40\textwidth,angle=0]{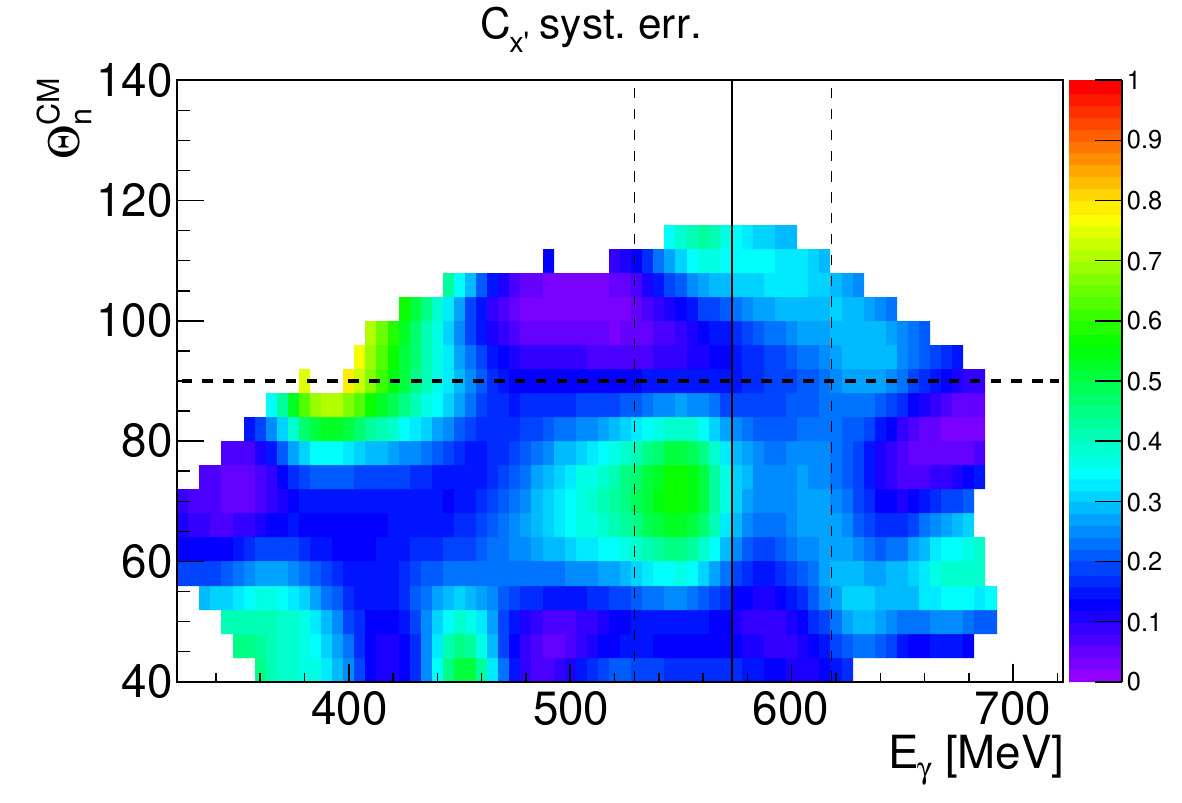}
\end{center}
\caption{(Colour online) (top) Two-dimensional $C_{x'}^n$ dependence as a function of neutron centre-of-mass angle, $\Theta^{CM}_{n}$ and photon energy, $E_{\gamma}$. The middle (top) plot shows the coresponding statistical (systematic) uncertainties. Contour lines of the sytematic uncertainties around 0.4 and 0.8 are also shown on the top and middle plots as dash-dotted lines. The vertical lines show the nominal $d^*$ pole position (solid) and width (dashed)~\cite{MBE1,MBE2}.}
\label{Cx2D}
\end{figure}

All systematic errors discussed above, and their kinematic dependencies, are combined in quadrature with the total systematic error for the complete polarimeter acceptance shown in Fig~\ref{Cx2D} (bottom). The typical magnitude of each of the systematic uncertainties are also summarised in Table~\ref{tab:sys}.
\begin{table}[ht]

\centering \protect\caption{Summary of systematic uncertainties}
\vspace{2mm}
{%
\begin{tabular}{|l|c|}
\hline
Type  &  error \\
\hline
\hline
$P_{\gamma}^{\odot}$    & 3\% \\
\hline
$A_y(^{12}$C$(n,p)X)$    & 10\%  \\
\hline
selection cuts    & 0.02 \\
\hline
$P_y$    & 0.02 \\
\hline
 amorphous/diamond   & variable$\sim 0.2$ \\
\hline
\end{tabular}} \label{tab:sys}
\end{table}

\begin{figure}[!h]
\begin{center}
\includegraphics[width=0.40\textwidth,angle=0]{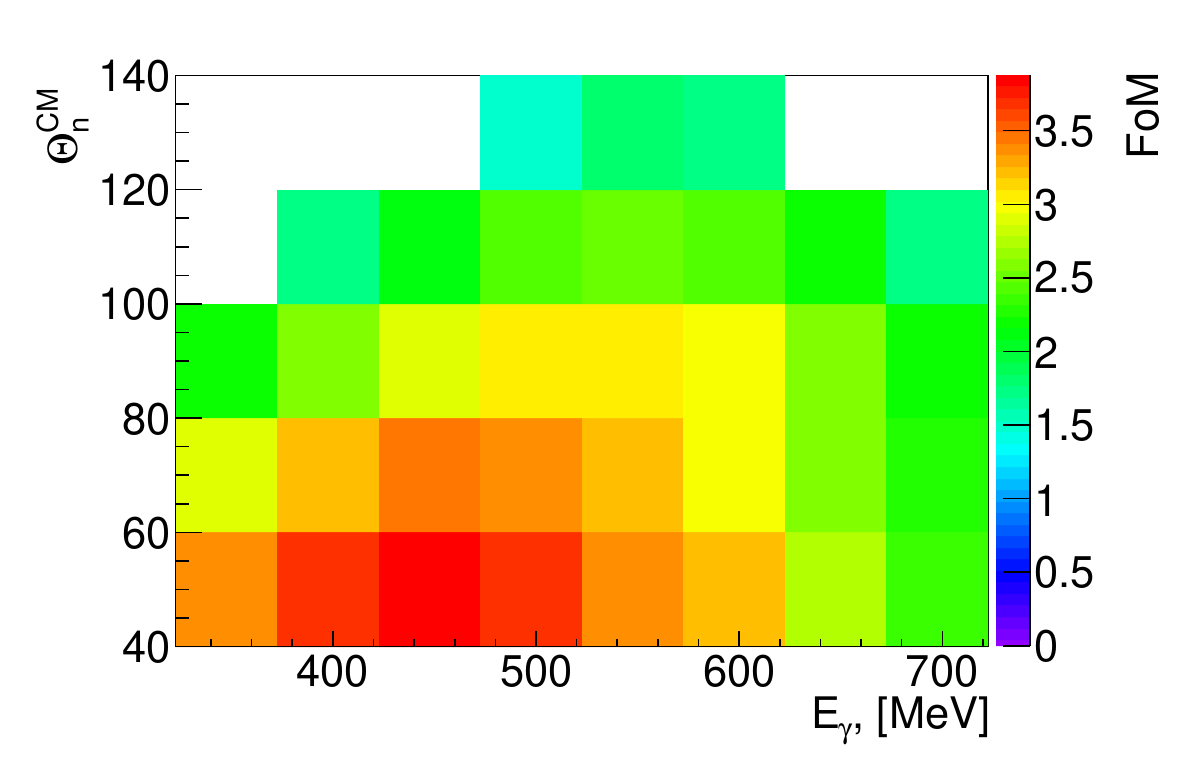}
\end{center}
\caption{Two-dimensional $C_{x'}^n$ sensitivity figure of merit dependence as a function of neutron centre-of-mass angle, $\Theta^{CM}_{n}$ and photon energy $E_{\gamma}$.}
\label{FoM}
\end{figure}

To give an overview of the polarimeter performance over the sampled parameter space of $E_{\gamma}$ and $\Theta_{n}^{CM}$, we constructed a figure of merit defined as $FoM=\sqrt{N}\cdot A_y\cdot P_\gamma^{\odot}$, where $N$ is the yield of events in each bin, $A_y$ is the average neutron analysing power in the bin and $P_\gamma^{\odot}$ is the circular photon polarisation. The FoM is shown in Fig.~\ref{FoM}. For much of the parameter space it is in the range of 3-4, reducing to around 1.5 near to the extremes of the acceptance.


As a further check on the analysis, a parallel $\gamma p\to n\pi^+$ analysis was performed on the same data. This reaction is well studied for this photon energy range and there is a general convergence of the predicted $C_{x'}$ between leading partial wave analysis groups (SAID~\cite{SAID_MA19} and BnGa~\cite{ZA19,BnGa1}). The obtained $C_{x'}$ distributions are in statistical and systematic agreement with both SAID and BnGa predictions over the measured range of photon energies and breakup angle \footnote{Note that BnGa and SAID adopt opposite sign conventions for this observable. In our work we adopt the same conventions as BnGa.)}~\cite{NpiPl}.

\section{Results}
Figure~\ref{Cx2D} (upper) shows the measured $C^n_{x'}$ over the full sampled phase-space, along with the associated statistical (middle) and systematic (bottom) uncertainties. The systematic errors represent a sum in quadrature of all the identified sources (see section 3). 

As expected, the obtained $C^n_{x'}$ data are consistent with the physical range within their statistical and systematic errors. For $E_{\gamma}\sim400-475$~MeV, close to the $\Delta$ resonance, localised regions of strongly negative $C^n_{x'}$ are observed for forward angle regions, while central angles of $\sim70^{\circ}-85^{\circ}$ show values closer to zero. There are indications of positive $C^n_{x'}$ at the most extreme angles, albeit associated with much larger systematic errors due to their proximity to the upper edge of the polarimeter acceptance  - see Fig.~\ref{Cx2D} (bottom). The location of the pole of the $d^{*}$ (extracted from elastic pn scattering~\cite{MBE1,MBE2}) is shown by the solid line on Figure~\ref{Cx2D} and its width indicated by the dashed lines. Rather rapid and continuous variation of $C^n_{x'}$ across the $d^{*}$ region is observed for $\sim\Theta^{CM}_{n}=55-85^{\circ}$ - ranging from close to zero to $\sim$-1 over the range of the $d^{*}$.
The large kinematic coverage achieved for this first determination of $C^n_{x'}$ will clearly provide valuable new constraints on our understanding of deuteron photodisintegration.

\begin{figure}[h]
\begin{center}
\includegraphics[width=0.40\textwidth,angle=0]{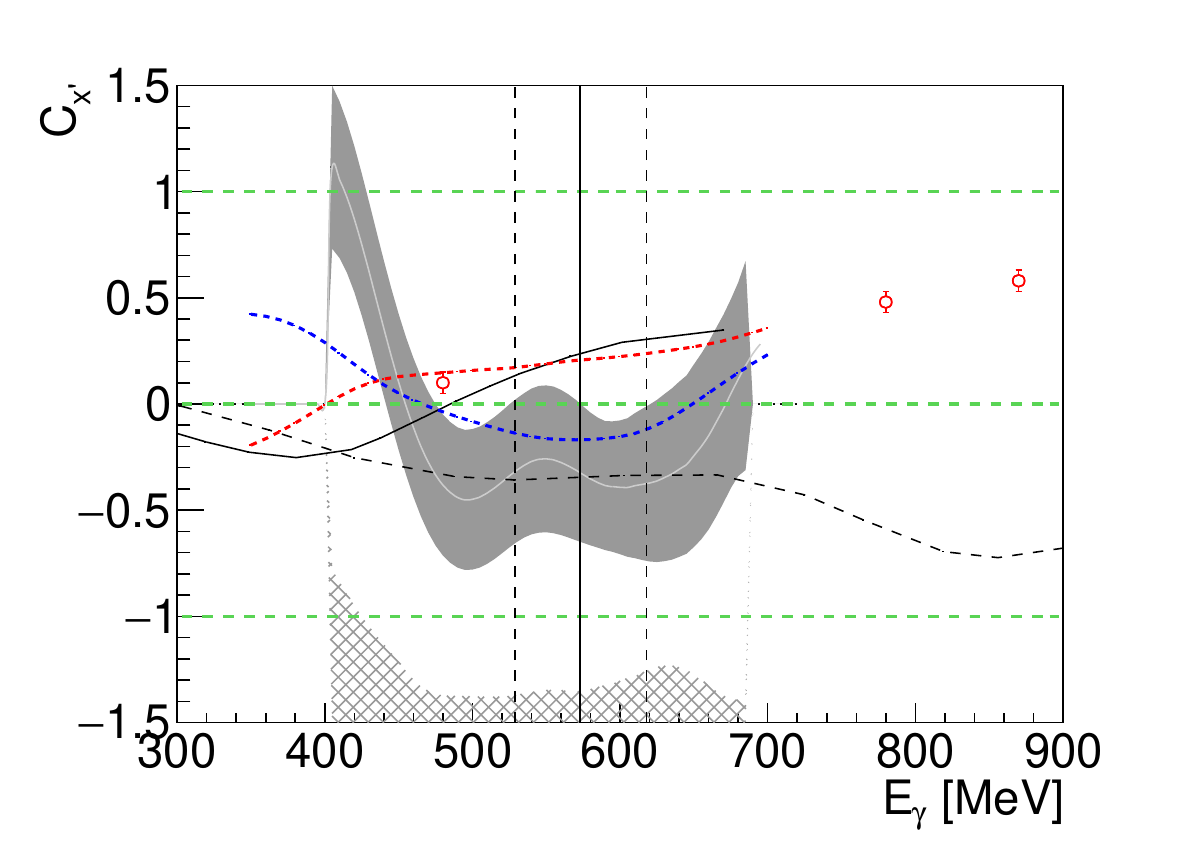}
\includegraphics[width=0.40\textwidth,angle=0]{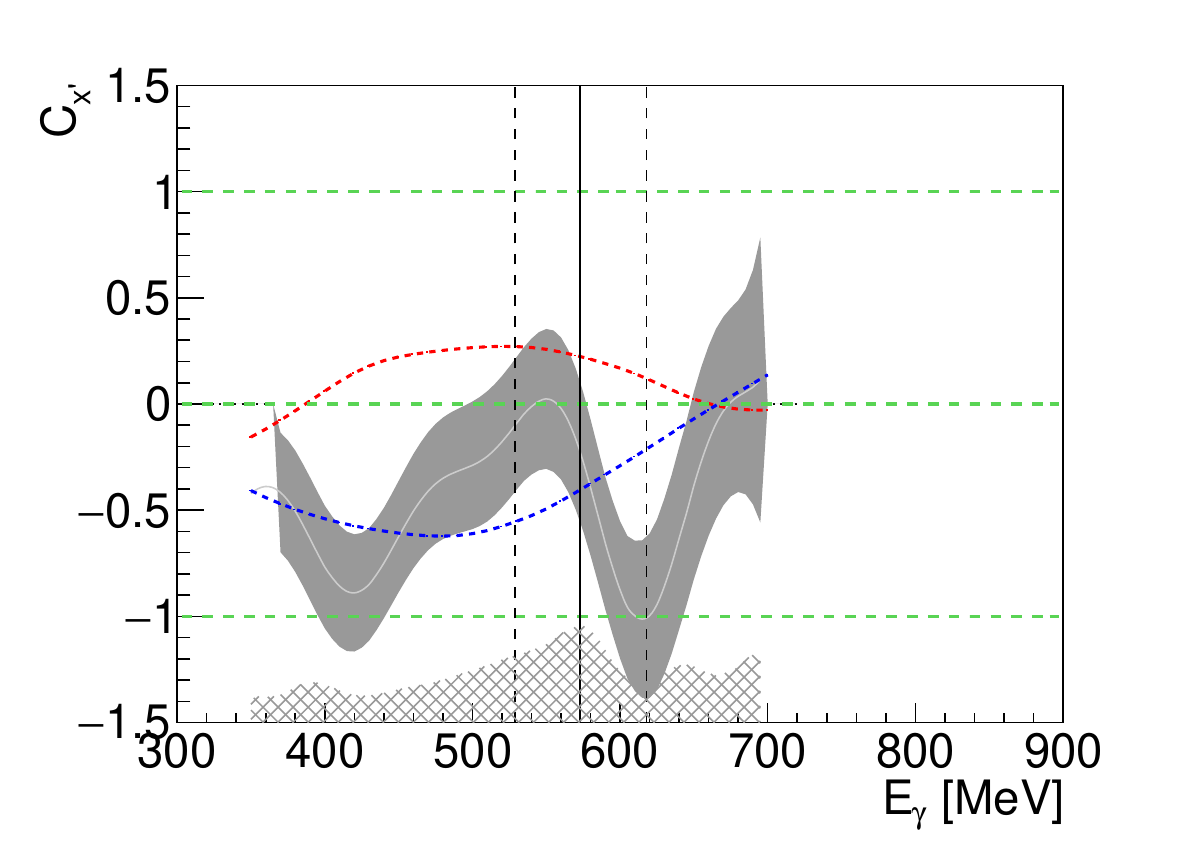}
\end{center}
\caption{$C^{n}_{x'}$ for the $\Theta_n^{CMS}=90^{\circ}$ (top) and $\Theta_n^{CMS}=60^{\circ}$ (bottom) are shown as a light grey line with statistical errors as a grey band and systematic errors as the hatched area on the bottom axis of each plot. Previous $C^{p}_{x'}$ from Ref.~\cite{JLabP} are shown as red markers. Calculations for $C^{p}_{x'}$ from Ref.~\cite{Aren} (Arenhoevel) and Ref.~\cite{Kang} (Kang) are shown as solid and dashed black lines respectively. Calculations for $C^{p}_{x'}$ (red) and $C^{n}_{x'}$ (blue) from Ref.~\cite{FixCx} (Fix) are shown as dotted lines. Vertical lines show nominal $d^*$ pole position (solid) and width (dashed).}
\label{Cx90}
\end{figure}

To explore the trends in $C^n_{x'}$ in more detail, in Fig.~\ref{Cx90}, we show $C_{x'}^n$ (light grey line) as a function of photon energy at a fixed angle  of $\Theta^{CM}_{n}=90^{\circ}$ (top) and $\Theta^{CM}_{n}=60^{\circ}$ (bottom). The statistical error bars are shown as a grey band and systematic errors as the hatched area on the bottom axis of each plot. As discussed above, the $\Theta^{CM}_{n}=90^{\circ}$ data suggest positive $C_{x'}^n$ for the lower photon energies in the $\Delta$ region below $\sim450$~MeV (albeit with large associated systematic errors) and negative values above this. For $\Theta^{CM}_{n}=60^{\circ}$, the data indicate generally negative  $C_{x'}^n$ with a minima around 420~MeV. inspection of Fig.~\ref{Cx2D} (upper) shows this minima would shift to higher photon energies for smaller $\Theta^{CM}_{n}$. Although providing first data for $C_{x'}^n$ in the $\Delta$ region, the current setup was not optimised for this photon energy range. Future measurements in the  $\Delta$ region with a more optimised setup, having higher beam polarisation (from employing a lower energy electron beam energy) and the use of thinner polarimeter material would provide data  over a wider kinematic range and with smaller systematic and statistical uncertainties.

The current work focuses on regions above the $\Delta$ to obtain the first scan of a double-polarisation observable ($C_{x'}^n$) through the region of the $d^{*}$. As $C_{x'}^n$ and the previously measured $P_{y'}^n$~\cite{MBPy} represent the real and imaginary components of the same combination of reaction amplitudes~\cite{GG}, then their correlation offers additional constraint on the properties of this amplitude combination. If the structure in $P_{y}^n$ centred around the $d^*(2380)$ hexaquark~\cite{MBPy} is indeed arising from its contribution to the imaginary part of the amplitude combination, then correlations with the real part (determined by $C_{x'}^n$) would provide new experimental constraint on the combination. For a single, isolated resonance the rapid variation in phase when crossing a resonance produces an s-shape in the real component of the resonant amplitude with a central value occurring at the pole. For the observable measured here, where the resonance occurs with backgrounds, the $C_{x'}^n$ and $P_{y'}^n$ observables represent the real and imaginary components of more than one amplitude. More detailed theoretical interpretation is clearly necessary before strong conclusions on any observed variations can be drawn. However, as the $d^*(2380)$ resonance is relatively narrow it is informative to assess any rapid variations in this region which could reflect its contribution,either directly or via interference with backgrounds.


With these caveats in interpretation, the photon energy dependence of $C_{x'}^n$ does indicate relatively rapid variation over the region of the $d^*(2380)$ (shown by the vertical dotted lines)in the  $\Theta_n^{CMS}=60^{\circ}$ bin. This bin may be expected to be more sensitive to $C_{x'}^n$ variations as it is centred on the node where $P_y$ is expected to vanish - assuming $P_y$ scales with the $P_1^3$ associated Legendre function, as discussed in ~\cite{MBPy}. Above the $d^*$ region, both angle bins indicate a rise in $C_{x'}^n$ towards positive values.

The $C_{x'}^n$ data are compared to the predictions of a theoretical model~\cite{FixCx} based on the diagrammatic approach used earlier by other authors, see e.g Laget~\cite{Laget}, Levchuk~\cite{Levchuk}. The model includes photocoupling to nucleon currents, meson exchange currents and isobar (resonance) contributions from $\Delta(1232)$, $P_{11}(1440)$, $D_{13}(1520)$ and $S_{11}(1535)$ resonances. The model incorporates all available $\gamma d\to pn$ data, including this $C_{x'}^n$ measurement. The resulting fits are shown as the dotted curves on Fig.~\ref{Cx90} for $C_{x'}^n$ (blue) and $C_{x'}^p$ (red). in the $\Delta$ region the model is consistent with the positive $C_{x'}^n$ data, albeit within the large systematic error of the data.  However above the $\Delta$ it is clear that this model does not predict any rapidly varying behaviour in the region of the $d^*$, despite the data being included in the fit. This suggests that nucleonic and resonance degrees of freedom in this region, as they are parameterised in the model, do not readily explain the variations in $C_{x'}^n$ observed for the $\Theta_n^{CMS}=60^{\circ}$ data. Clearly, further theoretical calculations which include the $d^*$ as a degree of freedom would be a critical next step  - along with more detailed theoretical treatment of the resonance contributions (we remark that the resonances included in the calculation include all relevant established resonances from the PDG). 
 
 It is informative to discuss the previous $C_{x'}^p$ data (red data points in Fig.~\ref{Cx90}) and its description by theory. There is a single datum at $\Theta^{CM}_{p}=90^{\circ}$ at $E_{\gamma}=475$~MeV which overlaps the current measurement. Our new $C_{x'}^n$ data is of comparable magnitude to this single $C_{x'}^p$ datum.  The Fix model~\cite{FixCx} (red dotted line) gives a good description of the single $C_{x'}^p$ datum. The coupled channel calculation of $C_{x'}^p$ from Arenhoevel~\cite{Aren} (black solid line; adopted from \cite{GG}) also shows agreement with the datum, although giving different predictions to the Fix model in the $\Delta$ region. The relativistic model of Kang~\cite{Kang} (black dashed line) predicts an opposite sign for $C_{x'}^p$ than evidenced in the data. It is interesting that above 700 MeV, the Kang predictions suggest a rapid variation in $C_{x'}^p$, compatible with regions where interference from the $N^*(1520)$ resonance may be expected (pole at 760~MeV width of 100~MeV). Unfortunately, both the Arenhoevel and Kang calculations were only published at this single breakup angle so no comparison for the $\Theta^{CM}_{p}=60^{\circ}$ bin can be made. The two $C_{x'}^p$ data points at higher $E_{\gamma}$ than the current data are also shown but are beyond the energy range of the available theories.

\section{\label{sec:final} Summary}

The neutron spin-transfer coefficient $C_{x'}^n$ in deuteron photodisintegration has been measured for $E_\gamma=370-700$~MeV and photon-deuteron centre-of-mass breakup angles for the proton of $40-120^{\circ}$, providing the first measurement of this fundamental observable.  At forward breakup angles a rapid and continuous variation is observed across the $d^{*}$ region. Comparison with a theoretical model based on a diagrammatic approach, fitted to all available deuterium photodisintegration data, and including all relevant (known) nucleon resonances did not reproduce the data in this region. The new data will provide a key challenge to future more detailed theoretical treatments of deuterium photodisintegration which include the $d^*(2380)$ as an explicit degree of freedom 

This new $C_{x'}^n$ data will be combined with future measurements of polarisation observables in deuteron photodisintegration from  polarised photon beams and a transversely polarised deuteron target at MAMI, progressing towards a first well-constrained partial wave analysis of the fundamental deuteron photodisintegration reaction and providing a benchmark dataset to challenge theoretical models. 
\section{Acknowledgements}
We are indebted to M. Zurek for providing us data on $n^{12}$C analysing powers. All data are available for downloading from PURE~\cite{PURE}. This work has been supported by the U.K. STFC (ST/V002570/1, ST/L00478X/2, ST/V001035/1, ST/P004385/2, ST/T002077/1, ST/L005824/1, 57071/1, 50727/1 ) grants, the Deutsche Forschungsgemeinschaft (SFB443, SFB/TR16, and SFB1044), DFG-RFBR (Grant No. 09-02-91330), Schweizerischer Nationalfonds (Contracts No. 200020-175807, No. 200020-156983, No. 132799, No. 121781, No. 117601), the U.S. Department of Energy (Offices of Science and Nuclear Physics, Awards No. DE-SC0014323, DEFG02-99-ER41110, No. DE-FG02-88ER40415, No. DEFG02-01-ER41194) and National Science Foundation (Grants NSF OISE-1358175; PHY-1039130, PHY-1714833, No. IIA-1358175), INFN (Italy), and NSERC of Canada (Grant No. FRN-SAPPJ2015-00023).

\end{document}